\documentclass{sig-alternate-2013}
\usepackage{url}
\usepackage{multirow}

\usepackage{array}
\usepackage{rotating}
\usepackage{color}
\usepackage{caption}
\usepackage{subcaption}
\newcommand{\CS}{\texttt{CS}}
\newcommand{\Physics}{\texttt{Physics}}

\permission{Copyright is held by the author/owner(s).}
\conferenceinfo{WWW'16 Companion,}{April 11--15, 2016, Montr\'eal, Qu\'ebec, Canada.} 
\copyrightetc{ACM \the\acmcopyr}
\crdata{978-1-4503-4144-8/16/04. \\
http://dx.doi.org/10.1145/2872518.2889375}

\begin{document}



\numberofauthors{1} 
\author{\alignauthor Dinesh Pradhan$^{1,a}$, Tanmoy Chakraborty$^{2,b}$, Saswata Pandit$^{3,a}$, Subrata Nandi$^{4,a}$ \\
\affaddr{$^a$Dept. of Computer Science \& Engg., National Institute of Technology, Durgapur, India} \\ \affaddr{ $^b$University of Maryland Institute for Advanced Computer Studies (UMIACS), College Park, MD 20742} \\
\email{\{$^{1}$dineshkrp,$^{3}$saswata.pandit94,$^{4}$subrata.nandi\}@gmail.com $^{2}$tanchak@umiacs.umd.edu}
} 

\title{On the Discovery of Success Trajectories of Authors}\vspace{-10mm}

\maketitle
\begin{abstract}
 
Understanding the
qualitative
patterns of research endeavor of scientific authors in terms of publication count  and their impact (citation) is important in order to quantify {\em success trajectories}. Here,
we examine the career profile of authors in computer science and physics domains and discover at least six different success trajectories in terms of normalized citation count in
longitudinal scale. 
Initial observations of individual trajectories lead us to characterize the authors in each category. 
We further leverage
this trajectory information to build a {\em two-stage stratification model} to predict future success of an
author at the early stage of her career. Our model outperforms the baseline with an average improvement of $15.68\%$ for both the datasets.

\end{abstract}
%
%
\vspace{-4mm}
\keywords{Success trajectories,; citation networks; prediction} 

\vspace{-3mm}
\section{Introduction}
An individual author's career trajectory is governed by a plenty of decisions and unforeseen events, that can
significantly impact her career.
As a result, the career trajectory is subjected to an author's past accomplishments and can
 be of different shapes in temporal scale.  
A \textit{success trajectory} can be defined with respect to different objectives, such as
research publications, funding, teaching ability etc. However, most important criterion accepted universally is
the {\em citation count} of an author's scientific publications.
Most of the author-centric indices, such as h-index, g-index captures either growth or saturation of research profiles, however fails to capture the decline of success. Analyzing the decline of success is similarly important to unfold several aspects, such as whether the authors are still active in research, how worthy are their recent publications, do they overcome the ``test of time'' challenge etc.

Here, we explore two massive datasets consisting of papers related to computer science and physics domains, 
and analyze the
success trajectory of authors in terms of the {\em normalized citation count} (ratio between total citations and total publications)  over the years. 
Interestingly, {\em we discover
at least six distinct categories of success trajectories, which to the best of our knowledge is revealed here for the first time in the
granularity of
individual authors}. 
Finally, we build a system which predicts (mean accuracy 15.68\% more than the baseline system) the future success of an author at the early stage of her career. 

\vspace{-4mm}
\section{Experimental Setup and Results}

\noindent{\bf Datasets.} We crawled two massive bibliographic datasets \cite{ChakrabortySNAM}: (i) ~\CS ~(2,473,171 articles in computer
science),
(ii)
~\Physics ~(425,399 articles in Physical Review journals).
%
After preprocessing, we consider 1,549,317
and 295,311
authors respectively from \CS ~and \Physics ~datasets whose publication informations are available for at least 10
years. 

\begin{figure*}[!t]
\centering
\begin{minipage}{.45\textwidth}
  \centering
  \includegraphics[width=0.95\linewidth]{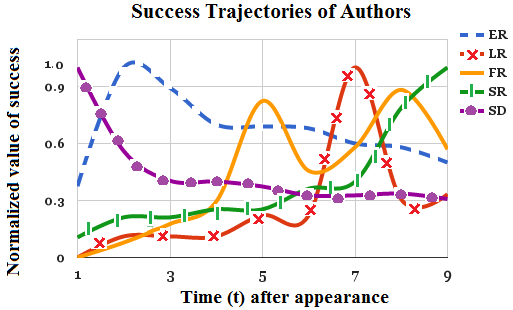}
 
\end{minipage}%
\begin{minipage}{.5\textwidth}
  \centering
  \includegraphics[width=\linewidth]{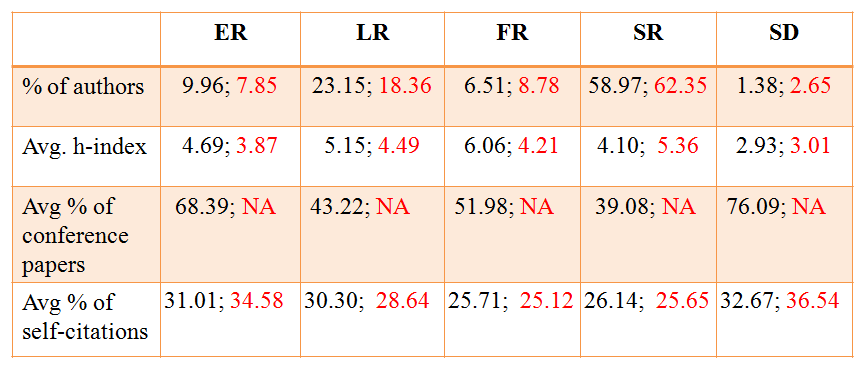}
\end{minipage}
\caption{(Color online)(Left) Sample examples (taken from \CS-dataset) of success trajectories; (Right) Characteristics of different
trajectory categories for \CS ~(black) and \Physics ~(red) datasets (NA: Not Applicable).}\label{fig}
\end{figure*}

\noindent {\bf Heuristics for trajectory discovery.}  To begin with, we take the selected sets of authors with the information of their papers and citations over time. An initial
three-year buffer window is provided to each author with the assumption that unlike for a paper, a few-years time frame is always
required for an
author
to set up her career. Therefore, we consider the fourth year of the career timeline of an author as the beginning of
the logical year of her career profile. Then we quantify the {\em success} of an author $a$ at year $t$ (termed as $S_{a}^{t}$)
by the ratio between the number of citations received by $a$ till $t$ (termed as $C_{a}^t$) and the number of papers published by $a$ till
$t$
(terms as $P_{a}^t$).
This is followed by a series
of data processing: firstly, to smoothen the longitudinal data points corresponding to an author, 
we use five-years moving average filtering for smoothing; secondly, we scale the data points by normalizing them with the maximum value present in the time series;
finally, we use two heuristics for peak detection: (i) the height of a peak should be at least 75\% of the
maximum peak-height, and (ii) two consecutive peaks
should be separated by more than 2 years; otherwise they are treated as a single peak.

\noindent{\bf Categories of success trajectories.} Remarkably, we observe six different patterns of success trajectories of authors based on 
the count and the position of peaks present in a trajectory (see Figure \ref{fig}(left)): (i) {\tt Early-risers} ({\tt ER}): authors whose
career peaks once within
initial $5$ years (but not in the first year) followed by a decay; (ii) {\tt Late-risers} ({\tt LR}): authors whose career
peaks once after at least 5 years since she has published her first paper, followed by a decay; (iii) {\tt Frequent-risers} ({\tt
FR}): authors whose
career peaks multiple times over the years; (iv) {\tt Steady-risers} ({\tt SR}): authors having a monotonic increasing growth of success
over the years; {\tt Steady-droppers} ({\tt SD}): authors whose career peaks in the first year followed by a monotonic decrease over the
years; and  {\tt Others} ({\tt OT}): apart from the above types, there exist a large volume of authors who on an average publish
less than one paper per year and receive less than one citation per year. Due to the lack of proper statistical evidences, we categorize
them into
a separate category.

\noindent{\bf Characterizing individual success trajectories.} Next, we attempt to understand the authors of individual categories in more
details (see
Figure \ref{fig}(right)). First, we calculate the percentage of authors in each category and observe that steady-risers are the major class
of authors, followed by late-risers; whereas steady-droppers are rare. Second, we measure the average impact of authors in each category
and notice that while in \CS ~domain frequent-risers are the most profound authors in terms of h-index, in \Physics ~steady-risers dominate
others, the reason being that physicists prefer publishing papers in Journals (see later).
However, as expected steady-droppers seem to be least prominent. 
Third, for \CS-dataset we notice that early-risers
and
steady-droppers tend to publish papers mostly in conferences, while steady- and frequent-risers prefer publications in journals. Forth and
most
interestingly, we observe that early-risers and steady-droppers  are mostly affected by self-citations
\footnote{{\tiny A citation is marked
as
self-citation if there is at least one author common in both citing and cited papers.}}. 
Had the self-citations been discarded from the
analysis, 53\% early-risers and 63\% steady-droppers have migrated to {\tt OT} category. 

A deeper inspection of the decay in the success trajectories of early-risers, late-risers and steady-droppers for \CS ~(\Physics) dataset
revels that around 82\% (79\%) cases the value of success drops due to the enormous volume of individual publications overshadowing the effect
of incoming citations. Further, we observe that during the time of decay, 46\% (37\%) of authors are
unable to retain their most prominent collaborators (in terms of h-index), indicating that the effect of collaboration might be a reason for
this decay. Interestingly, for both the datasets (\CS; \Physics) the rate of publications of steady-risers (2.06; 1.27) is least among
others (on average 4.32; 3.29), which indicates that formers tend to emphasize on {\em quality}, rather than quantity. 


\noindent{\bf Leveraging trajectory information for predicting success.} One crucial  question in this context is --  how can  the
trajectory information be leveraged for building real applications? Here we consider the task of predicting success (defined above) of an
author in future at the early stage ($t$ years after her first appearance) of her career. We consider the same set of
author-centric features (along with the first two years citations and publications of authors) and framework discussed by Chakraborty
et al. \cite{Chakraborty:2014} where Support Vector Regression (SVR) \cite{Chakraborty:2014} turned out to be the best learning framework.
We use 10-fold cross validation technique. The baseline is designed by training SVR on the {\em entire} training samples and predicting the
success of a query author by fitting the regression equation. On the other hand, we propose a {\em two-stage stratification learning} 
model \cite{HaroRS:07:SLD}. In stage one, a query author is mapped into one of the trajectories/strata\footnote{{\tiny Note that we
know the
category information of the authors present in the training set a priori, and therefore the training points are divided into six
categories.}} using a Support Vector Machine that learns from the same set of features used in the baseline. In the second
stage, {\em only} those authors corresponding to the category of the query authors are used to train the SVR module to predict the future
citation count of the query author. In this way, we remove the effect of random noise while training the regression model.
Experimental results show that our model achieves $15.09\%$ ($16.3\%$) and $14.7\%$ ($10.5\%$) more accuracy in term of mean squared error and
Pearson correlation coefficient respectively for \CS ~(\Physics) dataset\footnote{{\tiny The results are averaged over $t$,
ranging
from $3$ to $6$.}}.

\if
\noindent{\bf Modeling  the success trajectory of authors.} The crucial question one may ask in this context -- can we model the success trajectories? This in turn can be useful in predicting the success of an author at the early stage of her career. In particular, we attempt to design a new growth model of author-author citation network. We observe that the existing models \cite{} fail to substantiate these trajectories because these models are mostly built on the idea of ``preferential attachment'' and tend to ignore a crucial factor, i.e., ``aging''. Here we propose a new growth model that by constructing longitudinal author-author network would be able to infer these categorization of success trajectories. The model assumes that the number of authors, the time of their appearance in the network, and their publication      
\fi

\vspace{-4mm}
\section{Conclusions and Future work}
We discovered and characterized success trajectories of authors in two massive datasets. We believe that this information may be useful to develop models for performance prediction. Our study here for a span of at least initial $10$ years performance may be extended over several decades of an author's lifetime, which would lead to a complete characterization of her career. Understanding the dominant features among author's collaboration profile, affiliation, research domain,
etc. which primarily controls the success profile may
also be worth exploring further.


\vspace{-3mm}

\begin{thebibliography}{1}

\bibitem{Chakraborty:2014}
T.~Chakraborty, S.~Kumar, P.~Goyal, N.~Ganguly, and A.~Mukherjee.
\newblock Towards a stratified learning approach to predict future citation
  counts.
\newblock In {\em JCDL}, 2014.

\bibitem{ChakrabortySNAM}
T.~Chakraborty, S.~Sikdar, N.~Ganguly, and A.~Mukherjee.
\newblock Citation interactions among computer science fields: a quantitative
  route to the rise and fall of scientific research.
\newblock {\em SNAM}, 4(1):1--18, 2014.

\bibitem{HaroRS:07:SLD}
G.~Haro, G.~Randall, and G.~Sapiro.
\newblock {Stratification Learning: Detecting Mixed Density and Dimensionality
  in High Dimensional Point Clouds}.
\newblock In {\em NIPS}. 2007.

\end{thebibliography}

\end{document}